# Single quantum dot selection and tailor-made photonic device integration using nanoscale focus pinspot

Minho Choi[1], Mireu Lee[2,4], Sung-Yul L. Park[3], Byung Su Kim[1], Seongmoon Jun[1], Suk In Park[3], Jin Dong Song[3]*, Young-Ho Ko[2]*, Yong-Hoon Cho[1]*

[1]Department of Physics and KI for the NanoCentury, Korea Advanced Institute of Science and Technology (KAIST), Daejeon, Republic of Korea
[2]Quantum Technology Research Department, Electronics and Telecommunications Research Institute (ETRI), Daejeon, Republic of Korea
[3]Center for Opto-Electronic Materials and Devices Research, Korea Institute of Science and Technology (KIST), Seoul, Republic of Korea
[4]Department of Physics, Ulsan Nation Institute of Science and Technology (UNIST), Ulsan, Republic of Korea
*email: yhc@kaist.ac.kr; yhko@etri.re.kr; jdsong@kist.re.kr

Abstract

Among the diverse platforms of quantum light sources, epitaxially grown semiconductor quantum dots (QDs) are one of the most attractive workhorses for realizing various quantum photonic technologies owing to their outstanding brightness and scalability. There exist various material systems for these QDs based on their appropriate emission bandwidth; however, only a few material systems have successfully grown single or low-density QDs, which are essential for quantum light sources. In most other material systems, it is difficult to realize low-density QDs, and the mesa-etching process is usually undergone in order to reduce their density. Nevertheless, the etching process irreversibly destroys the medium near the QD, which is detrimental to in-plane device integration. In this study, we apply a nondestructive luminescence picking method termed as nanoscale focus pinspot (NFP) using helium ion microscopy to reduce the luminous QD density while retaining the surrounding medium. Given that the NFP can precisely manipulate the luminescence at nanoscale resolution, a photonic device can be deterministically fabricated on the target QD matched from both spatial and spectral points of view. After applying the NFP, we extract only a single QD emission out of the high-density ensemble QD emission. Moreover, the photonic structure of a circular Bragg reflector is deterministically integrated with the selected QD, and the extraction efficiency of the QD emission has been improved 27 times. Furthermore, this technique does not destroy the medium and only controls the luminescence. Hence, it is highly applicable to various photonic structures, including photonic waveguides or photonic crystal cavities regardless of their materials.

# Manuscript

Quantum light sources are essential building blocks for realizing quantum photonic technologies[1-6] and epitaxially grown semiconductor quantum dots (QDs) [7-31] are one of the most attractive platforms among various solid-state quantum light sources. These QDs are proving their potential as high-end quantum light sources with outstanding performance in terms of brightness, efficiency, purity, and indistinguishability as well as their integrability with optoelectronic devices.

The performance of these QDs can be drastically improved via photonic structure integration, and both spatial and spectral overlap between the QD and the optical mode of the photonic structure is essential[12-25]. However, the majority of QDs with cutting-edge performance are made by strain relaxation[13-28, 31] or droplet epitaxy[29, 30], which results in a random location and emission wavelength of the QD. Therefore, to deterministically integrate the QD with a photonic structure, the location and emission wavelength of the QD must be measured and identified first, and then the appropriately designed photonic structure can be fabricated accordingly.

Confocal photoluminescence (PL) mapping and wide-field PL imaging techniques are the most representative methods in which both spatial and spectral information of QDs are measured simultaneously at high resolution[20-24, 31, 32]. However, it is difficult to identify a single QD when the density of the QD is too high; for instance, if multiple QDs are excited by a single focused laser beam, optical signals from those QDs can be measured at the same time, and the information with respect to a single QD cannot be classified.

Hence, there have been a lot of efforts to form low-density QDs uniformly during the growth process, and there has been some success in a few material systems, including GaAs QDs grown by droplet epitaxy[33]. However, the formation of low-density QDs based on well-known Stranski-Krastanov (S-K) growth mode, such as In(GaAl)As on GaAs, III-N[34], and III-Sb[35] QDs, remains a challenging issue. Although researchers intentionally create a density gradient along the wafer, the selected area of low-density part is limited[28]. Thus, it is difficult to expect homogenous properties from the QDs across the whole wafer, and the optical characterization and density estimation processes of the QD become more complicated. Moreover, low-density site-controlled QDs can be grown by the nanoscale patterning[36-39] or local heating[40] on the wafer. However, the optical quality of these QDs are far short of the QDs naturally formed QDs after strain relaxation.

Hence, if the density of the QDs can be reduced after growing them and a single QD can be picked at the predetermined position, then it can lead to a sensational breakthrough for utilizing these high-density QDs from diverse material systems. Nanoscale patterning followed by mesa etching has been widely used to reduce QD density and separate a single QD[22-24, 32]. However, a critical issue arises for the mesa etching process, in which integration with an on-chip photonic structure, i.e., photonic waveguide, photonic crystal cavity, and circular Bragg reflector (CBR), is no longer possible because all the materials surrounding the QD have already been etched.

**Nanoscale focus pinspot (NFP) technique**

We developed a nondestructive nanoscale luminescence picking method utilizing focused-ion-beam (FIB) induced luminescence quenching and termed it as the nanoscale focus pinspot (NFP)[8]. Here, we use helium-ion beam due to its high directionality inside the crystal structure[41]. Given that we can selectively quench unwanted emissions without destroying the photonic structure, this method can be an optimal solution for high-density QDs to reduce their luminous QD density and still enable them to integrate with photonic structures in the lateral direction.

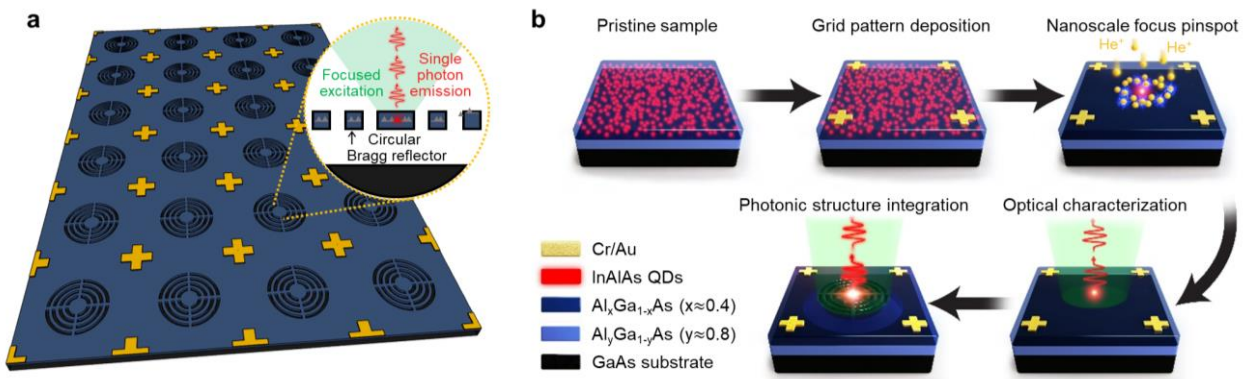

**Fig. 1| Concept of the deterministic integration of the selected QD with photonic structure. a**, Schematics of site-selective single QD integrated with CBR structure. **b**, Procedures of a single QD selection from the ensemble QDs using NFP and deterministic fabrication with the CBR structure.

In this study, we successfully reduced the luminous QD density from 40 $\mu m^{-2}$ to 1 $\mu m^{-2}$ using the NFP technique and observed only a single QD emission among the ensemble QDs. The selected single QD location can be predetermined with respect to the metallic grid pattern on the surface, and the QD can be deterministically integrated with photonic structures as shown in **Fig. 1a**. We first measured the optical properties (i.e., emission wavelength and polarization) of the selected QD and then designed and fabricated a CBR accordingly. The extraction efficiency of the exciton emission from the identical QD was improved 27 times after integration with the CBR structure.

**Fig. 1b** shows a schematic of the entire process. A minimal reproducible density of self-assembled InAlAs/AlGaAs QD sample was grown via molecular beam epitaxy in the modified S-K growth mode[42]. The density of the QDs was estimated at approximately 40 $\mu m^{-2}$ via atomic force microscopy of the top QD layer (**Supplementary Information 1**). From around this density or higher, the formation of QDs are stable and reproducible. A metallic grid pattern is formed on the sample surface via the lift-off process. Therefore, subsequent experiments of FIB irradiation, optical characterization, and CBR structure fabrication can be conducted according to the grid pattern. Given the FIB can image the grid pattern and perform the in-situ NFP process, the reliability of the NFP process is approximately a few nanometers in scale. The most important advantage of this procedure is that the dimensions of the photonic structure can be determined with respect to the optical properties (i.e., wavelength and polarization direction) of the selected QD after the NFP process. Therefore, the spectral and spatial overlap between the QD and photonic structures can be determined deterministically.

**Single QD selection using NFP**

During the NFP process, we used helium ion microscopy (HIM), which has a higher quenching resolution than the conventional FIB using gallium ions. It is important to note that scanning electron microscopy (SEM) is not installed inside the HIM equipment, so it is difficult to image the sample with a helium-ion beam without causing luminescence quenching owing to ion bombardment. Therefore, we imaged only the area of the grid pattern and performed the NFP process accordingly as shown in **Fig. 2a**. For the NFP process, we irradiated the helium-ion beam in a doughnut-shaped pattern with an outer diameter ($d_{out}$) of 20 µm and varied the inner diameter ($d_{in}$) from 0.5 µm to 5.0 µm. We used $1.0 \times 10^{14}$ cm$^{-2}$ helium ion irradiation dose condition to fully quench the QD emission where the HIM was irradiated (**Supplementary Information 2**). At this dose, we observed luminescence quenching followed by a doughnut-shaped irradiation pattern by measuring the low-temperature cathodoluminescence (CL) image (**Supplementary Information 3**). It should be noted that there were no imprints on the surface from the SEM image. Hence, we can integrate tailor-made photonic circuits for the chosen QD after measuring the optical properties of the QD after the NFP process because there is no structural destruction.

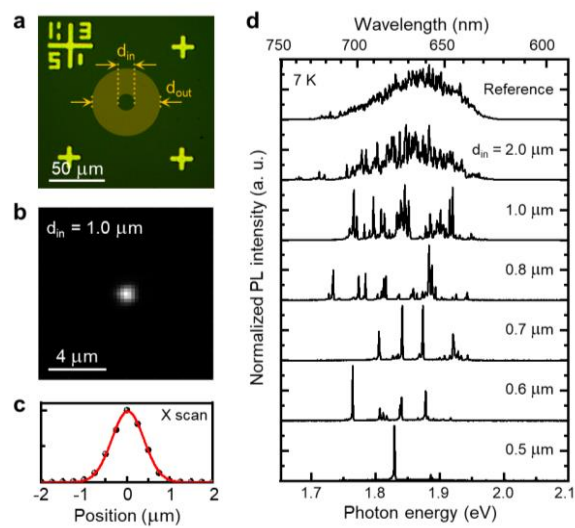

**Fig. 2| NFP process for a high-density ensemble QDs. a**, Optical microscopy image of the metallic grid pattern and a schematic of a doughnut-shaped helium ion irradiation. **b**, Low-temperature micro-PL mapping of the

NFP-processed QD emission. **c**, Line-scan of the micro-PL mapping. **d**, Inner diameter dependent micro-PL spectra of the QD emission after the NFP process.

The optical properties of the luminous QDs at the center of the doughnut-shaped irradiation pattern were measured via low-temperature PL experiments at 7 K. We used a continuous-wave laser with a wavelength of 532 nm and an objective lens with a numerical aperture of 0.65 and magnification power of 50X for the optical measurement. **Fig. 2b** shows the low-temperature PL mapping results of the QD emission where the NFP process was performed with an inner diameter of 1.0 μm. The mapping image represents a spatially resolved QD emission intensity, where all PL intensities of the QD emission from 640 to 720 nm were integrated. Based on the mapping image, we can observe the spotlighted QD emission after the NFP process, and the full width at half maximum (FWHM) of the bright spot is approximately 0.750 ± 0.005 μm (**Fig. 2c**), which is similar to the excitation laser beam spot. The PL mapping image of the QD layer without the NFP process is shown in **Supplementary Information 4**, where QD emission is observed throughout. **Fig. 2d** shows the micro-PL spectra of the bright spot after the NFP process as the inner diameter of the doughnut-shaped irradiation pattern is varied. When the inner diameter of the irradiation pattern decreases from 5.0 to 0.5 μm, the number of QD emission peaks decreases. Eventually, we obtained only a single QD peak after the NFP process when the inner diameter of the doughnut-shaped pattern was approximately 0.5 μm. We can adjust the luminous region, not only a single emission spot, by simply adjusting the irradiation pattern. Moreover, the process can be done quite fast thanks to the extremely low dose condition for luminescence quenching, less than 10 second is taken for 100 μm$^2$. Hence, highly scalable integration with multiple QDs can be expected.

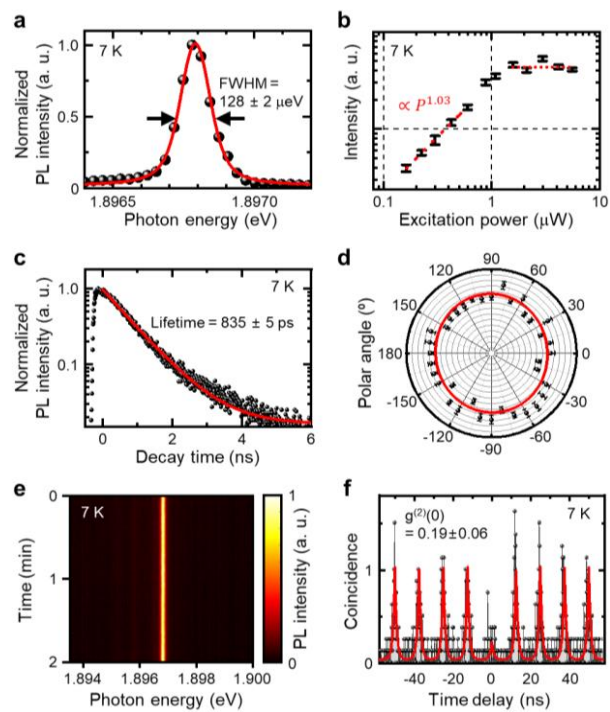

**Fig. 3| Optical characterization of a selected single QD after the NFP process. a**, High-resolution micro-PL spectrum of a single QD emission. **b**, Power-dependent PL intensity of a single QD emission. **c**, Time-resolved PL intensity of a single QD emission. **d**, Polarization dependence of a single QD emission. **e**, Long timescale spectral diffusion of a single QD emission. **f**, Second-order photon correlation of a single QD emission.

**Fig. 3** shows the optical characterization of the selected single QD emission after the NFP process. Here, we used a pulsed laser with a wavelength of 532 nm using the second-harmonic generation of a Ti:sapphire laser pumped optical parametric oscillator system. **Fig. 3a** shows the highly resolved QD emission spectrum, where FWHM is approximately $128 \pm 2$ μeV from the Voigt-functional fitting. Since we can only measure the ensemble QDs emission before the NFP process, we are not able to compare the linewidth of the identical QD emission before and after the NPF process. However, in a previous study, we have shown that the linewidth of the QD emission does not change after the NFP process from the site-controlled single QD emission[8].

**Fig. 3b** shows the power-dependent PL intensity of the QD emission. Specifically, the PL intensity represents the integrated QD emission spectra after Voigt-functional fitting. The power dependence follows the relationship $I(P) = I_{sat} \times (1 - A \times \exp(-P/P_{sat}))$. We can clearly observe a one-to-one proportionality of the exciton emission at the lower excitation power region, and the emission intensity is saturated at the high excitation power region, $I_{sat}$, which corresponds to the brightness of the QD. **Fig. 3c** shows the time-resolved PL signal of the single QD emission, and the lifetime is obtained as $835 \pm 5$ ps from the exponential decay functional fitting, $I(t) = I_0 \exp(-t/\tau)$, where τ denotes the lifetime. **Fig. 3d** shows the linear polarization dependence of the QD emission, and we can observe that there is no significant degree of linear polarization of the QD. It is well known that the degree of linear polarization of the QD emission is closely related to the geometrical and strain anisotropy of the QD. Hence, it is plausible to claim that the QD has symmetric geometry and strain along its lateral direction.

**Fig. 3e** shows the spectral diffusion which indicates the coherence property of the quantum emission over a longer timescale compared to that of the linewidth. We measure the QD emission spectra 120 times with 1 s for each spectrum. We can clearly see that there is no blinking, and from the Voight-functional fitting of every 120 spectra, we can extract the standard deviation of the peak position and integrated intensity of the QD emission. As a result, the standard deviation of the emission energy was approximately 2.0 μeV over 120 s, which is negligible compared to the spectral resolution of the measurement setup of approximately 100 μeV. Also, the standard deviation of the integrated intensity of the QD emission is approximately 1.8%. Hence, we can conclude that there is no degradation in the coherence properties of the QD emission after the NFP process in terms of short and long timescale.

**Fig. 3f** shows the second-order photon correlation of the QD emission. Under pulsed-laser excitation, the correlation follows the formula $g^{(2)}(t) = \sum A_i \exp(-|t - t_i|/\tau)$, where t denotes the time delay and τ denotes the effective recombination time. The autocorrelation of the QD emission at zero time delay, $g^{(2)}(t=0)$, is approximately 0.19 ± 0.06. A value below 0.5 is evidence of a single-photon source. As shown in **Fig. 2d**, we selected a single QD emission from the ensemble QD emission after the NFP process by measuring the spectrum. Here, we can confirm that the single QD emission is truly a single photon emission by measuring its autocorrelation. The single-photon purity of the QD emission is greater than 0.8 after the NFP process, and the remaining multiphoton events are less than 0.2. The measured single-photon purity can be further improved by optimizing the growth process (i.e., the absence of a wetting layer)[29] and optical characterization process (i.e., resonant excitation)[13].

**Quantum efficiency of the selected QD**

We successfully selected a single QD emission from high-density QDs and discussed its coherence properties. In **Fig. 4**, we focus on the quantum efficiency of ensemble QDs and a single QD after the NFP process. Given that the NFP process selectively quenched the luminescence from undesired areas, there might be an influence of surrounding defects on the selected emitter. Therefore, we compared the temperature-dependent PL intensity and lifetime of the QD emissions before and after the NFP process. Before the NFP process, it was not possible to identify a single QD emission; therefore, we measured the ensemble QD emission at three different positions and statistically analyzed the results (**Fig. 4a–c**). Furthermore, we identified a single QD emission after the NFP process; therefore, temperature-dependent PL intensity and lifetime were measured for a selected single QD after the NFP process of the doughnut irradiation with an inner diameter of 0.7 μm (**Fig. 4d–f**). Additionally, we selected a few QD emissions using the mesa-etching process and measured the temperature-dependent PL (**Supplementary Information 5**) to compare the results with the NFP-processed QD.

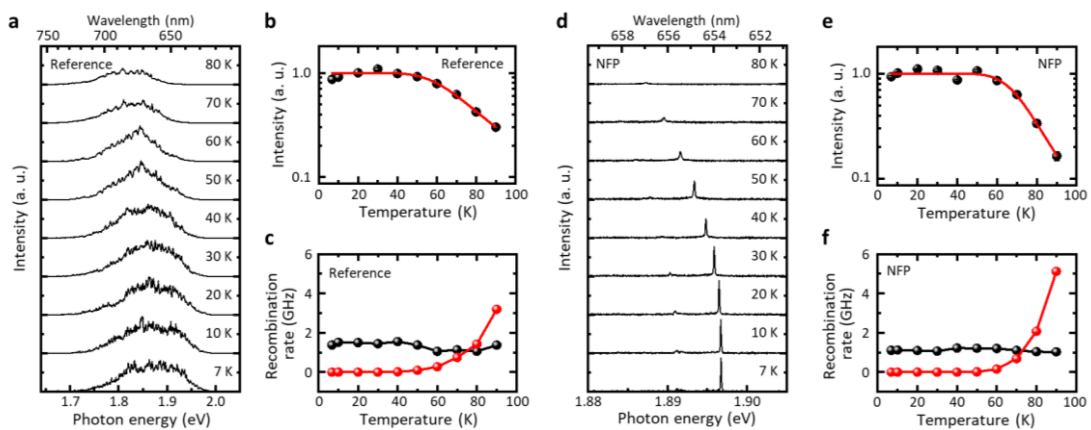

**Fig. 4| Temperature-dependent optical characterization of ensemble QDs and a single QD after the NFP process. a**, micro-PL spectra of ensemble QDs emission without the NFP process. **b**, Temperature-dependent PL intensity of ensemble QDs emission without the NFP process. **c**, Temperature-dependent radiative (black) and nonradiative (red) recombination rate of ensemble QDs emission without the NFP process. **d**, micro-PL spectra of a selected single QD emission after the NFP process. **e**, Temperature-dependent PL intensity of a selected

single QD emission after the NFP process. **f**, Temperature-dependent radiative (black) and nonradiative (red) recombination rate of a selected single QD emission after the NFP process.

**Fig. 4a and 4d** show the temperature-dependent micro-PL spectra of the ensemble QDs and single QD emissions before and after the NFP process, respectively. Correspondingly, **Fig. 4b** shows the integrated PL intensity of all the ensemble QD emissions depending on their temperature, whereas **Fig. 4e** shows the temperature dependence of the single QD emission intensity after the NFP. The results strongly follow the Arrhenius plot, $I(T) = I_0/(1 + A \times \exp(-E/k_B T))$, where E, $k_B$, and T denote the escape energy to the nonradiative channel, Boltzmann constant, and temperature, respectively. Based on the fitting, we can define and extract the effective quantum efficiency ($QE_{eff}$) as the relative emission intensity ratio of the QD between 70 K and 7 K, $QE_{eff} = I_{70K}/I_{7K}$, and compare the values of the pristine ensemble QDs and single QD after the NFP and mesa-etching processes. The $QE_{eff}$ value of the pristine ensemble QDs is approximately 0.60 while the values of the single QD after the NFP and mesa-etching processes are 0.62 and 0.04, respectively. We can clearly observe that the quantum efficiency of the selected QD after the NFP process is significantly higher than that of the QD after the mesa-etching process. The large difference between the NFP and mesa-etching process is plausibly due to the absence of etched surface states after NFP[43]. Conversely, the $QE_{eff}$ values of the single QD after the NFP process are almost similar to those of the pristine ensemble QDs; therefore, we can conclude that the NFP process does not affect the quantum efficiency of the selected QD emission, which is similar to the coherence property discussed in **Fig. 3**.

To further investigate the carrier dynamics inside the single QD after the NFP process, we measured the time-resolved PL at different temperatures from 7 to 90 K (**Fig. 4c and 4f**). We extracted radiative and nonradiative recombination rates from the measured lifetime using the following formula: $k_{measured}(T) = k_{rad}(T) + k_{nonrad}(T)$; $k_{measured} = 1/\tau_{measured}$; $QE = k_{rad}/(k_{rad} + k_{nonrad})$, where $k_{measured}$, $k_{rad}$, and $k_{nonrad}$ denote the measured, radiative, and nonradiative recombination rates, $\tau_{measured}$ denotes the measured lifetime, and QE denotes the quantum efficiency obtained from the temperature-dependent PL intensity (**Fig. 4b and 4e**). Based on this formula, we can successfully separate the radiative and nonradiative recombination rates from the measured recombination rates at each temperature. The results show that the radiative recombination rate is almost constant at varying temperatures, which is a characteristic of 3-dimensionally-confined QD emission, whereas the nonradiative recombination rate increases at elevated temperatures. Nonradiative recombination became dominant from 80 K for both non-treated reference QDs and single QD after the NFP process, which implies that there was no degradation of the QDs after the NFP process in terms of quantum efficiency and radiative carrier recombination process.

**Deterministic integration of the selected QD with photonic structure**

We showed that the NFP process enables the selection of a luminous single QD from high-density QDs and does not degrade its coherence properties and quantum efficiency. Above all those advantages, structural non-destructiveness is the most important aspect of the NFP process wherein on-chip integration is still available unlike the mesa etching process. Here, we demonstrate the deterministic integration of a selected single QD with a CBR following the procedure above (**Fig. 1b**). Here, the sample has AlGaAs sacrificial layer of high-Al contents, we can selectively wet etch the layer and form a free-standing structure (**Supplementary Information 6**). Given that we already know the location of the QD with respect to the grid markers, we can realize a spatial overlap between the QD and photonic structure. Additionally, the optical properties of the selected QD after the NFP process can be measured first, followed by the design and fabrication of the tailor-made photonic structure for the selected QD can be performed afterward; therefore, the spectral overlap between the QD and photonic structure can be realized (**Supplementary Information 7**).

Specifically, we selected the CBR structure for integration with the selected QD after NFP. The CBR structure was adopted for various quantum emitters to improve their extraction efficiency[44, 45]. In most cases, these QDs are surrounded by a high-refractive-index medium; therefore, a large portion of the emission cannot escape the host medium and is difficult to be collected by free-space optical systems. Hence, we measured the brightness of the identical single QD emission before and after CBR structure integration and determined the improvements of the extraction efficiency after integration. CBR is fabricated by the following process: after the NFP process and optical characterization, aligned electron-beam lithography with respect to the metallic grid patterns followed by reactive ion etching with chlorine-based gases, and the wet etching undercut process using HF solution provides a freestanding CBR structure that contains the selected QD at the center.

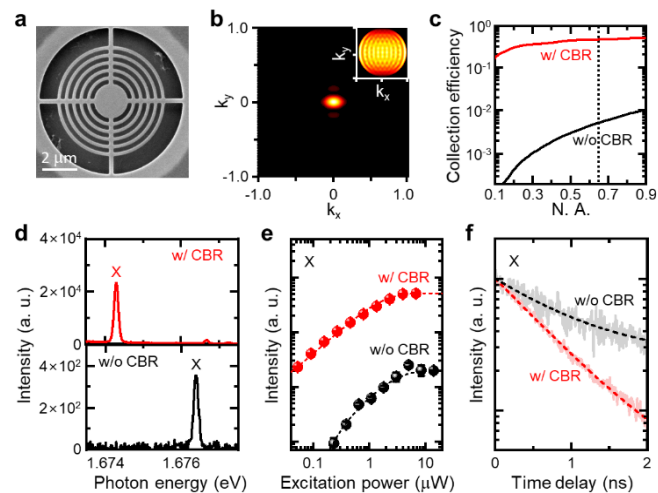

**Fig. 5| Deterministic integration between a selected single QD after the NFP process and a CBR. a**, SEM image of a CBR. **b**, Optical simulation of a far-field irradiation pattern of a single QD in a CBR structure. The inset image denotes the far-field irradiation pattern of a single QD in a slab structure. **c**, Calculated extraction efficiency of a single QD in a CBR structure and a slab structure. **d**, Low-temperature micro-PL spectra of an identical single QD emission before and after the CBR integration. Note that the y-axis scales are different for two cases. **e**, Power-

dependent PL intensity of exciton (X) emission from the identical single QD selected by the NFP process before and after the CBR integration. **f**, Time-resolved PL intensity of exciton (X) emission from the identical single QD selected by the NFP process before and after the CBR integration.

**Fig. 5a** shows an SEM image of the freestanding AlGaAs CBR containing InAlAs QDs. **Fig. 5b** shows the simulated far-field irradiation pattern of the QD coupled with the CBR structure, while the inset image shows the result from the QD before CBR fabrication. Highly directional Gaussian-like irradiation from the QD coupled with the CBR structure can be observed, which clearly differs from the QD inside a planar structure. **Fig. 5c** shows the simulated extraction efficiency of the QD emission with respect to the numerical aperture (NA) of the collecting optics. The calculated extraction efficiency of the QD coupled with the CBR structure is approximately 0.4 with a collecting system of NA 0.65 that we used. This can be doubled when we collect upward and downward emissions by implementing a reflector at the bottom, i.e., a distributed Bragg reflector. For comparison, the calculated extraction efficiency of a single QD in a slab structure before fabricating the CBR structure was approximately 0.005. Therefore, the extraction efficiency of the QD emission was enhanced by approximately 80 times with CBR coupling.

**Fig. 5d** shows the micro-PL spectra of the selected single QD after the NFP process with and without CBR structure integration under the same excitation condition with a 532-nm pulsed laser at 7 K. By comparing the optical properties of an identical QD before and after the CBR structure, it is straightforward to discuss about the extraction efficiency enhancement after CBR structure integration. Here, we precisely verify and calibrate the optical measurement setup between two measurements of before and after CBR structure fabrication by measuring the reference QD sample (**Supplementary Information 8**). Therefore, we can clearly conclude that the intensity of exciton (X) emission is significantly improved owing to coupling with the CBR structure (**Fig. 5d**). The peak position of the QD coupled with the CBR structure undergoes a redshift, which is common for other QDs (**Supplementary Information 9**), and it is expected to be due to the strain change on the QD with the freestanding structure. **Fig. 5e** shows the integrated PL intensity of the QD emission before and after the CBR structure integration. From fitting with the following function, $I(P) = I_{sat} \times (1 - A \times \exp(-P/P_{sat}))$, the saturated intensity $I_{sat}$, which corresponds to the brightness of the QD emission, was enhanced by approximately 27 times (from 198 to 5,290). The enhancement factor is of the same order as the calculated value of 80, whereas some difference may be originated from an imperfection of designed fabrication. **Fig. 5f** shows the time-resolved PL intensity of the QD emission before and after the CBR structure integration. From the exponential fitting, the recombination lifetime of the exciton before and after the integration is about 940 and 720 ps, respectively. Here, the QD emission is measured under low-temperature, and the QD is far apart from the etched surface, therefore we can consider the measured lifetime as a radiative term. Hence we can conclude that there is a radiative recombination rate enhancement, so called Purcell enhancement, with a factor of 1.3. Eventually, the tailor-made CBR structure for high extraction efficiency has been successfully made.

**Conclusion**

To sum up, we developed the effective and scalable method for building low-density and site-controlled quantum emitters by utilizing FIB-induced luminescence quenching. This NFP process enables single-QD selection from high-density ensemble QD emission without degrading its coherence properties and quantum efficiency. Most importantly, the NFP process does not destroy its surrounding medium and only selectively quenches its luminescence site. Therefore, on-chip integration includes not only a CBR, but also waveguides and photonic crystal cavities available after the NFP process. In this study, we demonstrated the integration between an NFP-processed QD and a CBR as a tailor-designed photonic structure. Moreover, when the size of the optical interaction between the QD and circuit becomes larger, the meaningfulness of the NFP process increases. For instance, monolithic on-chip integration of QDs can be a great candidate for on-chip quantum photonic processors. There is a critical challenge regarding the unwanted emission or absorption from other QDs or wetting layers spread throughout the wafer. To avoid this issue, hybrid integration has been adopted, in which passive materials, such as silicon or silicon nitride, are used for large-area photonic circuits. However, the fabrication difficulties and low coupling efficiency of these hybrid integrations remain critical problems. In our work, by using the NFP process, we can select a single luminous QD from the multiple QD emissions; hence, it is plausible to solve these issues that arise from these monolithic quantum photonic platforms.

## Methods

### Quantum dot growth

The samples were grown in a solid source MBE system (model Comapct21) equipted with ion getter pump and cryogenic pump supplied by Riber company. Semi-insulating undoped (001) 2 inch GaAs substrates were introduced into the MBE chamber and heated up to 600 °C for the surface oxide removals under arsenic dimmer ambient of beam equivalent pressure of $8.6\times10^{-6}$ Torr. The growth temperature was decreased to 580 °C for the growth of 400 nm-thick GaAs buffer layers. On it, a 1000 nm-thick $Al_{0.8}Ga_{0.2}As$, a 4nm-thick GaAs and a 60 nm-thick $Al_{0.4}Ga_{0.6}As$ layers were grown successively at 600 °C. Then, InAlAs QDs were formed with modified S-K growth method at 530 °C[42]. That is, 3 repetion of 4 s-InAlAs, 20 s-growth interruption, 10 s-InAs and 20 s-growth interruption. The formation of 3-dimensional pattern was obserbed in the 3$^{rd}$ repeation of QD formation. During QD formation, the rotation of manipulator was stopped for the gradation of QD density across the substrate. Finally, a 60 nm-thick $Al_{0.4}Ga_{0.6}As$ was grown. The growth rate of $Al_{0.8}Ga_{0.2}As$ and $Al_{0.4}Ga_{0.6}As$ were 0.15 nm/s, 0.09 nm/s, respectively. The growth rate of InAs is 0.04 nm/s. The growth interrpuption stands for the close of the all K-cell shutters.

### Nanoscale focus pinspot

Commercialized helium ion microscopy (ZEISS, Orion NanoFab) is used for NFP process. The kinetic momentum of the accelerated He-ion is smaller than the other type of focused ion beams, i.e. Ga-ion, Ne-ion, Si-ion, under the same acceleration voltage which results in the greater luminescence quenching resolution. Given that the collision cross-sectional area is minimum for helium ion (there is no H-ion FIB), the accelerated ions can propagate further inside the sample and quench the luminescence in greater depth. In this study, we use an acceleration voltage of 30 kV, beam current of 1–2 pA, dwell time of 10 μsec, and pitch size of 10 nm. The resultant of the dose condition is about $10^{14}$ ions/cm$^2$ which completely quenches the InAlAs QDs luminescence (Supplementary Information 2).

### Optical measurements

We use either a continuous-wave and pulsed laser of 532 nm wavelength to optically excite the QD. For continuous-wave laser, we use a diode-pumped solid-state laser with about a few μW power. For pulsed laser, we use a Ti:sapphire (Coherent, Chameleon-Ultra II) laser of 800 nm wavelength to pump the optical parameteric oscillator (Coherent, Chameleon Compact OPO) which results in the pulsed laser of 1064 nm wavelength. Second harmonic generation from the external nonlinear crystal generates the pulsed laser of 532 nm wavelength. The QD sample is cooled by a closed-loop liquid helium cryostat (Montana Instrument, Cryostation C2). We use a objective lens (Mitutoyo) with a magnification factor of 50 and numerical aperture of 0.65. We filtered the reflected laser beam from the emission signal by using dichroic beamsplitter (Semrock, Di03-R532-t1). Spectrum of the QD emission is measured by a monochromator (Princeton Instrument, HR750) and charge-coupled device (Princeton Instrument, PIXIS 400). Both time-resolved photoluminescence and second-order photon colleration are measured by using avalanche photodiode (ID Quantique, ID100) and time-correlated single photon counting (Picoquant, FluoTime 200).

**Photonic structure fabrication**

Cr/Au metallic grid markers are made upon the QD sample by the photo-lithography followed by electron beam evaporation and liftoff process. NFP process can be done at determined position with in-situ nanoscale imaging of those premade markers. We can extract the emission wavelength information of those selected QDs after the NFP process. Tailor-made CBR structures are designed with optical simulation according to those emission wavelengths of selected QDs. Given that we already know where the NFP process is done, we can deterministically fabricate the appropriate CBR structure at that specific position by aligned electron beam lithography followed by the inductively-coupled plasma reactive ion etching with chlorin-based gases. Then, the freestanding CBR structure was formed by selectively etching the sacrificial layer of $Al_{0.8}Ga_{0.2}As$ layer with HF solution under the $Al_{0.4}Ga_{0.6}As$ layer containing the selected QD. Here, the 4 nm-thick GaAs etch stop layer between the $Al_{0.8}Ga_{0.2}As$ and $Al_{0.4}Ga_{0.6}As$ layer prevents the damage for the $Al_{0.4}Ga_{0.6}As$ slab during the wet etching process

**Data availability**

The data that support the findings of this study are available from the corresponding authors upon reasonable request.

**Acknowledgements**

The authors appreciate KAIST Analysis center for Research Advancement and DGIST CCRF for supporting experiments. This work was supported by the the National Research Foundation (NRF-2022R1A2B5B03002560 and NRF-2020M3E4A1080112) and Institute of Information & Communications Technology Planning & Evaluation (IITP-2020-0-00841) of the Korean government, and the Samsung Science and Technology Foundation under Project Number SSTF-BA1602-05.

**Author contributions**

Y.-H.C. and M.C. initiated the study and designed all experiments. M.C., B.S.K. and S.J. performed optical characterizations. M.L. and Y.-H.K. designed and fabricated the circular Bragg reflector. S.-Y.P., S.I.P. and J.D.S. provided the quantum dot sample. Y.-H.C., J.S and Y.-H.K. conceived and supervised the project. M.C. and Y.-H.C. wrote the manuscript, supported by all co-authors.

**Competing interests**

The authors declare no competing interests.